\documentclass[twocolumn,usenatbib,prl,floats,usegraphicx]{revtex4}
\usepackage{epsfig}
\usepackage{graphicx}
\usepackage{graphics}
\usepackage{amsfonts} 
\usepackage{amsmath, amssymb, stmaryrd}  
\usepackage{psfrag}  
\def\be{\begin{equation}}
\def\ee{\end{equation}}
\def\bi{\begin{itemize}}
\def\ei{\end{itemize}}

\begin{document}
\title{Shape of Clusters as a Probe of Screening Mechanisms in Modified Gravity}
\author{Claudio Llinares and David F. Mota}
\affiliation{Institute of Theoretical Astrophysics, University of Oslo, N-0315 Oslo, Norway}
\begin{abstract}
Scalar fields are crucial components in high energy physics and extensions of General Relativity.   The fact they are not observed in the solar system may be due to a mechanism which screens their presence in high dense regions.  We show how observations of the ellipticity of galaxy clusters can discriminate between models with and without scalar fields and even between different screening mechanisms.  Using nowadays X-ray observations we put novel constraints on the different models.
\end{abstract}
\maketitle
General Relativity (GR) is a successfully tested theory in solar system scales and below.  Assuming this theory is valid also at cosmological scales gave rise to the $\Lambda$CDM model, which has its foundations in two unknown components: dark matter and dark energy.  The nature of these two components could therefore be an indication of the breaking down of Einstein's gravity on large scales.  This  has motivated the proposal of several theories which modify GR at astrophysical scales \citep[][]{2012PhR...513....1C}.

An imperative requirement to all Modified Gravity proposals is they all must recover GR in the solar system. This is done via a screening mechanism.
Presently, there are three main screening mechanisms: Vainshtein \citep{vain}, Symmetron \citep[][]{{2010PhRvL.104w1301H}} and Chameleon \citep[][]{{cham}}.
We focus on the two later ones since they are described by scalar degrees of freedom, and they have the common feature of emerging at the onset of nonlinear structure formation.

The key feature of screening mechanism is to switch off the extra degrees of freedom inside matter overdensites (small scales), and to switch them on in the cosmological background (large scales).  When the scalar fields are {\it{on}} a fifth force emerges between the matter particles. When it is {\it{off}} (the field is screened) the fifth force disappears and GR is recovered.
This is a highly nonlinear process, since the scalar fields are strongly coupled to matter and have highly nonlinear bare potentials. 

The aim of this Letter is to investigate signatures that the chameleon and the symmetron fields imprint in the formed nonlinear structures, which can be measured by nowadays' experiments, and so be used as probes of extensions of GR and the required screening mechanism to reproduce it within the solar system.

Dark matter halos are not spherical, and its density scales differently along the $x$, $y$ or $z$ directions. Such anisotropic shape of the halos (ellipticity), leads to an anisotropic screening mechanism of the scalar fields.  The result is that the fifth force between dark matter particles can be present in one direction, while being almost nonexistent in another. This anisotropy may lead to an increase in the ellipticity of the clusters, which can be measured by lensing or X-ray observations, and so be  used as tracers of the inherent screening mechanism.

It is well known that the Newtonian potential of triaxial systems acquires a shape that is more spherical than the matter-density distribution itself (see for instance results from simulations in \citet[][]{2011ApJ...734...93L}).  In the case of strongly coupled scalar fields their iso-surfaces are expected to follow closely the matter-iso-density contours. Due to the screening,  the fifth force range and couplings change along the matter-iso-density contours, leading to modifications in the shape of the system. In this Letter we  test this conjecture by studying the 3D distribution of scalar fields that correspond to triaxial dark matter halos, and use X-ray observations to put bounds on the models  with and without scalar fields and to distinguish between different screening mechanisms.  

We present calculations of the scalar fields and Newtonian potential for a fixed density distribution. Therefore, our study will not account for the time evolution of the system, which can be seen as a restriction in our results.  Nevertheless, one has to take into account that the X-ray component of relaxed clusters is in hydrostatic equilibrium and thus, their shape follows that of the total gravitational potential (GR + modifications).  By making our calculations for two different shapes of the underling dark matter distribution, we show that whatever is the effect of modified gravity in the system while virialized, the relative difference with respect to GR is not very sensitive to the shape of the underlying DM distribution. Therefore, the result is not expected to be sensitive to the time evolution.  This must be confirmed with cosmological simulations, which go beyond this work.

The Newtonian potential $\phi_N$ is given by:
\be
\nabla^2 \phi_N = \frac{3}{2}\frac{\Omega_m H_0^2}{a} \delta, 
\ee
where $\delta$ is the over-density defined as $\delta\rho/\rho_b$, $\rho_b$ is the mean density of the universe, $\phi_N$ is the perturbation in the metric, $\Omega_m$ is the mean density of the universe in terms of the critical density, $H_0$ is the Hubble constant and $a$ is the expansion factor.

The symmetron model \citep[][]{2010PhRvL.104w1301H}  is defined by the following effective potential:
\be
V_{s,eff}(\phi_s) = \frac{1}{2} \left(\frac{\rho}{M^2} - \mu^2 \right) \phi_s^2 + \frac{1}{4}\lambda \phi_s^4, 
\ee
which leads to the following equation of motion in the static limit:
\be
\nabla^2 \phi_s = a^2 \left[-\mu^2 \phi_s + \lambda \phi_s^3 + \frac{1}{M^2}\rho\phi_s \right], 
\ee
where $\mu$ and $M$ are mass scales, $\rho$ is the matter density and $\lambda$ is a length scale.  We normalize the field $\phi_s$ with the minimum of the potential $\phi_{s,0}$ that corresponds to zero density and  is given by:
$\phi_{s,0}^2 = \frac{\mu^2}{\lambda}$.
By dividing the whole equation by $\phi_{s,0}$, defining the dimensionless quantity
$\chi_s=\frac{\phi_s}{\phi_{s,0}}$,
and taking into account that 
$\rho_{SSB} = M^2 \mu^2$ at  $z_{SSB}$,
we get:
\be
\label{poisson}
\nabla^2 \chi_s = \frac{a^2}{2 \lambda_{s,0}^2} \left[ -\chi_s + \frac{\eta\chi_s}{a^3(1+z_{SSB})^3}  + \chi_s^3 \right].
\ee
where $\eta$ is the matter density field normalized with the mean density of the universe, and $\lambda_{s,0} = \frac{1}{\sqrt{2} \mu}$ is the range for the field that corresponds to zero density.

The associated effective potential for the chameleon  is:
\be
V_{c, eff}(\phi_c) = M_c^{4+n} \phi_c^{-n} + \rho e^{\beta\phi_c/M_{pl}}, 
\ee
where $M_c$ has units of mass, $\beta$ is dimensionless, $n$ is a positive constant and $M_{pl}$ is the Planck mass.  The linearized equation of motion for the scalar field is:
\be
\nabla^2\phi_c = -\frac{n M_c^{4+n}}{\phi_c^{n+1}} + \frac{\beta}{M_{pl}} \rho.
\ee
As in the symmetron model and for numerical convenience, we normalize $\phi_c$ with the minimum of the effective potential $\phi_{c,0}$.  In the chameleon case, the minimum diverges when the density goes to zero.  Thus, in this case we normalize with the minimum that corresponds to the mean density of the universe.  After including the range of the field for this particular density
\be
\lambda_c^2 = \left(\frac{d V_{c.eff}}{d\phi_c}\right)^{-1} = \left( n (n+1)M_c^{4+n}\phi_{c,0}^{-(n+2)} \right)^{-1}
\ee
 and re-normalizing the field $\chi_c=\phi_c/\phi_{c,0}$, we get
\be
\nabla\cdot \left[ q \omega_c^{q-1}\nabla\omega_c  \right] = \frac{1}{(n+1)\lambda_c^2}\left[ \eta - \frac{1}{\omega_c^{q(n+1)}} \right], 
\ee
where $\chi_c = \omega_c^q$ is chosen to facilitate the numerics.

To solve the field equations we use a Fourier based method for the Newtonian case and an implicit multigrid non-linear solver for both scalar field equations.  The code uses a uniform grid and is an extension of  \citep[][]{llinares_thesis}, to which we added both scalar field solvers.  The boundary conditions are periodic.  The three solvers were tested successfully against analytic solutions for a sphere of uniform density located in the center of the box.

To test the possibility that the presence of a scalar field can have an impact in the shape of clusters, we calculated the Newtonian potential and both scalar fields for a density distribution given by a NFW profile \citep[][]{{1997ApJ...490..493N}}.  The virial radius $R_v$ of the object was fixed to 1 Mpc/h, which corresponds to an object of $10^{14}$ M$_{\odot}$ (i.e. a cluster of galaxies).  Following mass-concentration relations coming from simulations \citep[]{2011MNRAS.411..584M}, we choose a concentration of 6.3 for our halo.  The density distribution was defined in every node of the grid following the analytic profile.  The size of the box is 16 Mpc/h and $512$ nodes per dimension were employed, which corresponds to a spatial resolution of about 30 kpc/h.  The underlying cosmology needed to normalize the density profile was chosen as $\Lambda$CDM, defined by $\Omega_m=0.3$ and $H_0=70$ km/sec/Mpc.  All our analysis are at redshift $z=0$.

As we want to measure how closely the scalar field follows the triaxiality of the density distribution, we need to assume a density profile.  We choose the NFW profile:
\be
\frac{\rho(r)}{\rho_0} = \frac{\delta_{\mathrm{char}}}{\Omega_0} \frac{1}{(r/r_s)(1+r/r_s)^2}, 
\ee
but using an ellipsoidal radius $k = \sqrt{x^2+\frac{y^2}{q^2} + \frac{z^2}{s^2}}$ instead of the radius $r$ that corresponds to spherical coordinates.  The axial ratios of the density distribution where fixed to be $(q^2,s^2)=(0.5,0.3)$.  

Our calculations show that the chameleon model is sensitive to the behavior of the density far from the center of the halo.  In order to get stable results, we immersed the halo in a background with a constant density of 0.4 the mean density of the universe.  We choose a value lower than the mean density of the universe to take into account the fact that clusters are surrounded by voids, and thus immersed in under-dense regions.  In fact, we find that the Newtonian and symmetron values are independent of the presence of this background. 
\begin{figure}
  \begin{center}
    \includegraphics[width=0.3\textwidth]{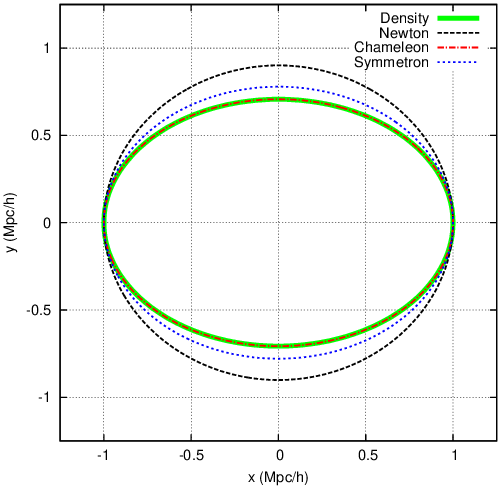}
    \caption{Contours of density distribution, Newtonian potential and symmetron and chameleon fields for the triaxial halo in the plane $x-y$.  The contours where chosen such that all of them pass through the point $(R_v,0,0)$.  Here $z_{SSB}=1.6$ and $\lambda_{s,0}=\lambda_c=1.1$.} 
    \label{fig:contours}
  \end{center}
\end{figure}
\begin{figure*}[!t]
  \begin{center}
    \includegraphics[width=0.6\textwidth]{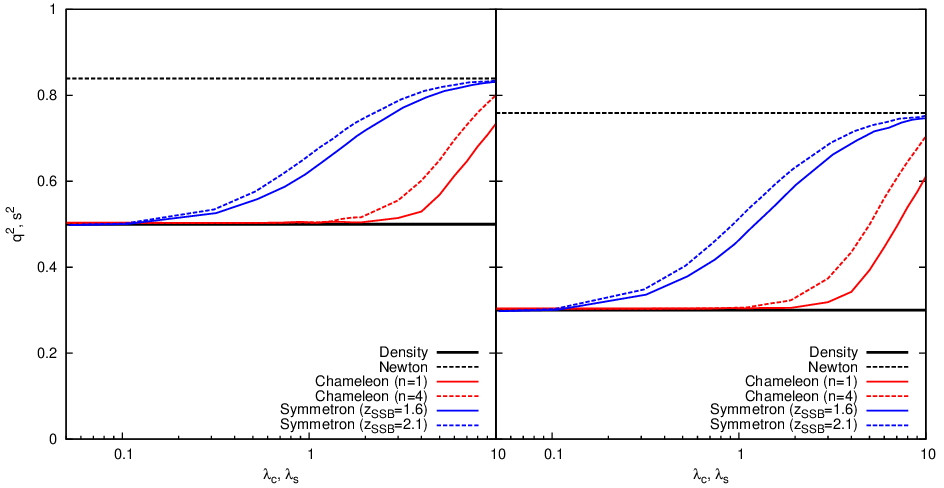}
    \caption{Axial ratios $q^2$ (left) and $s^2$ (right) for the Newtonian potential and the two implementations of the scalar field as a function of the range of the scalar field.  The two curves for the symmetron model correspond to two different values of $z_{SSB}$.  In the case of the chameleon, we show results for two different values of $n$.  The continuous black lines are the values that correspond to the density distribution and the dotted black lines the ones that correspond to the Newtonian potential.} 
    \label{fig:axial_both}
  \end{center}
\end{figure*}

Given the solutions of the field equations, we now measure the shape of their iso-surfaces:  the axial ratios of the iso-surfaces of a given distribution $f$ (in our case the Newtonian potential and both scalar fields) is:
\be
M_{ij} = \int f x_i x_j d^3x.
\label{definition_shape}
\ee
As the absolute value of the scalar fields does not decrease with radius, we defined $f$ as a renormalized version of the fields.  These iso-surfaces can be approximated by ellipsoids, described by the ellipsoidal radius $k$ with axial ratios $q^2=\frac{M_{xx}}{M_{zz}}$ and $s^2=\frac{M_{yy}}{M_{zz}}$, where $M_{xx}$, $M_{yy}$ and $M_{zz}$ are the eigenvalues of $M_{ij}$.  The integral in that equation is computed by suming over the grid up to twice the virial radius.  As in the symmetron model the scalar field can be  screened up to values of $k$ of the order of $2 R_v$, we extended the domain of the integral for the most extreme cases (with very small values of $\lambda_{s,0}$ and $z_{SSB}$).  Thus, in the symmetron case, we integrated up to the maximum between $2 R_v$ and the radius $k$ at which $f$ reaches 0.5.  In all the cases, the summation was made up to no more than $3 R_v$.  The shape of the region in which the integral is computed was obtained iteratively as in\citep[][]{1991ApJ...378..496D}.

Fig.\ref{fig:contours} shows iso-densities and iso-surfaces for the Newtonian potential and both scalar fields in the plane that corresponds to the major and intermediate axis of the system.  As expected, the Newtonian iso-potential falls apart from the matter-iso-densities and acquire a form that is much closer to sphericity.  In the scalar field cases, we find not only that their iso-surfaces follow the density distribution much closer than the Newtonian potential, but also that there are differences between them (even when the range of the field is the same for both models).  

We find that symmetron fields tend to be more spherical than the chameleons. The reason for such difference lies in the mechanism driving the screening of the fifth force: in the symmetron there is a defined threshold density above which the field decouples from matter (the fifth force vanishes), while in the chameleon the fifth force disappears slowly and continuously (its range becomes shorter) as the density becomes higher and higher.

To quantify the differences between the models and understand the dependence of the result with the model parameters, we calculate the axial ratios for every model.  The values obtained for the Newtonian potential are $q^2=0.84$ and $s^2=0.76$, which differ considerably from the input parameters given for the density.  The results obtained for the symmetron and chameleon models are shown in Fig.\ref{fig:axial_both}.  To make a fear comparison we show here the results as a function of the range of the field $\lambda_s$ and $\lambda_c$ that corresponds to the mean density of the universe in both cases.  We find a different behavior for each scalar field model.  The symmetron mechanism tends to give more spherical iso-surfaces when increasing both $\lambda_s$ and $z_{SSB}$.  The chameleon mechanism is insensitive to changes in $\lambda_c$ up to ranges that are larger than the virial radius itself. Once again, this reflects the difference in the way the fields are screened.

To test the stability of our calculations, we made resolution and box size convergence tests by increasing the resolution by a factor of two while keeping the box size constant and also by increasing both box size and resolution by the same factor.  The tests were made for model parameters that are representative of iso-surfaces with small and large values of $q^2$ and $s^2$.  We find that our solutions are independent of the resolution.  The Newtonian and symmetron solutions are stable with respect to changes in the box size.  In the case of the chameleon model, we find that the solution is much more sensitive to the distance between the halo and the boundary. Nevertheless, the actual change in the axial ratios when going from 16 to 32 Mpc/h in the box size is only of the order of 5\%, which is far from the variations we see when changing from model to model.  In any case, the environment around non-isolated clusters is expected to change the solutions, but not the bulk of the signal.

To test the sensitivity of our results when the underlying density distribution is changed, we repeated our analysis with a less extreme model taken from simulations \citep[][]{2006ApJ...646..815S}: $(q^2,s^2)=(0.6241,0.459684)$.  We find that our results (the relative difference between the shape of the scalar field and that of the density) are rather insensitive to the underling density distribution.

An estimation of the importance of the effect from the observational point of view can be made under the assumption that the X-ray component follows the iso-surfaces of the total potential (Newtonian plus scalar field). The ellipticity $\epsilon=1-b/a$ of the projection on the sky of these iso-surfaces should then be the same as of the gas density distribution.  This assumption makes also possible to obtain constraints on the model parameters, since there is a set of parameters for which the difference between the ellipticity that we predict by assuming standard gravity and by including the fifth force is larger than the errors in present X-ray observations.  \citet{2012arXiv1201.2168L} reported measurements of ellipticities using Chandra and ROSAT observations originally presented in \citet[]{2009ApJ...692.1033V}.  They found almost constant ellipticities in the radial range of  $0.05\lesssim r/r_{500} \lesssim 1$ which are given by $\epsilon \approx 0.18 \pm 0.05$.  We refer the reader to \citet{2012arXiv1201.2168L} and references therein for details on these observations.

\begin{figure}[!t]
  \begin{center}
    \includegraphics[width=0.47\textwidth]{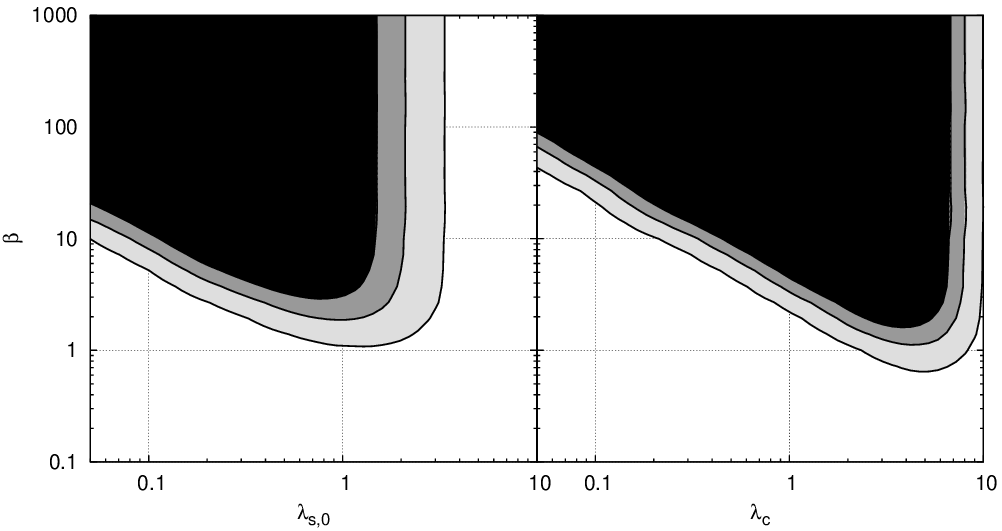}
  \end{center}
  \caption{Contours of relative difference $\Delta\epsilon/\epsilon_N$ between ellipticities that correspond to the modified models (LEFT: symmetron, RIGHT: chameleon) with respect to Newtonian gravity. At small $\lambda$ the screening mechanism emerge and the allowed parameter space increases. Here $z_{SSB}=2.1$ and $n=4$.}
    \label{fig:param_space}
\end{figure}

The calculation of the total potential was made by taking into account that the geodesics equation at redshift $z=0$ has the following form in the symmetron case:
\be
\ddot{\textbf{x}} + 2 H \dot{\textbf{x}} + \nabla\left(\phi_N + 6H_0^2 \Omega_m \lambda_{s,0}^2 \frac{(1+z_{SSB})^3}{2}\beta^2 \chi_s^2 \right) = 0 
\nonumber
\ee
and can be written for the chameleon as:
\be
\ddot{\textbf{x}} + 2 H \dot{\textbf{x}} + \nabla\left(\phi_N + 6H_0^2 \Omega_m  \lambda_c^2 \frac{(n+1)}{2} \beta^2 \chi_c \right) = 0.
\nonumber
\ee
We estimate the ellipticity of these iso-surfaces of total potential by taking the mean value over random projections.  Fig.\ref{fig:param_space} shows contours of the relative difference $\Delta\epsilon/\epsilon_N$ between ellipticities that correspond to the total potential associated to the modified models and to the Newtonian potential.  The regions from black to light grey correspond to values larger than three, two and one times $\sigma_{obs}/\epsilon_{obs}$.  In other words, the black region is ruled out with more than $3\sigma_{obs}/\epsilon_{obs}$.  When making this comparison, we assume that observations and Newtonian theory give the same values for the ellipticity.  By taking into account a possible bias of the predictions of the standard model towards more spherical halos \citep{2012arXiv1201.2168L}, one can relax slightly the constraints and include models with higher values of $\beta$ that are excluded here.

From Fig.\ref{fig:param_space},  it is clear that for high couplings and small ranges, the chameleon model is less constrained than the symmetron. In other words, the chameleon tends to give more spherical objects.  That seems to be in tension with Fig.\ref{fig:axial_both}: in there, the symmetron  field tends to be more spherical.  This can be understood taking into account the dependence on $z_{SSB}$ in the symmetron geodesics equation.  This extra factor makes the total potential to have a stronger contribution from the scalar field and thus, a larger ellipticity than the chameleon, even in the case that its intrinsic distribution is more spherical.

In summary, we propose an astrophysical test which can be used as a probe to detect or differentiate screening mechanisms associated to scalar fields which are present in gravity theories which modified GR at scales larger than the solar system.  We show that the existence of such screening mechanism can strongly affect the shape of galaxy clusters. 
Starting from a dark matter density distribution that corresponds to a non-spherical cluster of galaxies, we measure the shape of the iso-surfaces that correspond to Newtonian potential and two scalar field models (symmetron and chameleon).  We find that both scalar field models give iso-surfaces that follow much more closely the density distribution than the Newtonian potential.  Furthermore, we find that the shape of the iso-surfaces also depends on the mechanism used to screen the fifth force: the symmetron model tends to give more spherical distributions than the chameleon one. Since, present observations show some tension between the shape of real clusters \citep[]{{2012arXiv1201.2168L},{2012MNRAS.420.3213O}} and predictions obtained from simulations \citep[][]{2011ApJ...734...93L},
 our results indicate that if scalar fields make any difference, it is in the right direction to correct the discrepancy in the observations. %
Finally, using recent data from  X-ray observations, we   calculate novel constraints on the coupling $\beta$, and the ranges, $\lambda_{s,0}$ and $\lambda_c$, of the scalar fields' fifth force.
\acknowledgments{
DFM and CLL thank funding from Research Council of Norway, and H. Winther and H. Dahle for discussions.}
\bibliography{references}
\end{document}